\documentclass[12pt,preprint]{aastex}

\newcommand{\fluxunit}{ergs cm$^{-2}$ s$^{-1}$}

\shortauthors{Wang et al.}

\begin{document}

\title{Chandra X-ray Sources in the LALA Cetus Field\altaffilmark{1}}

\author{J. X. Wang\altaffilmark{2}, Z. Y. Zheng\altaffilmark{2}, 
S. Malhotra\altaffilmark{3}, S. L. Finkelstein\altaffilmark{4},
J. E. Rhoads\altaffilmark{3}, C. A. Norman\altaffilmark{5},
T. M. Heckman\altaffilmark{5}
}
\begin{abstract}
The 174 ks $Chandra$ Advanced CCD Imaging Spectrometer exposure 
of the Large Area Lyman Alpha Survey (LALA) Cetus field is the second of 
the two deep $Chandra$ images on LALA fields.
In this paper we present the $Chandra$ X-ray sources detected in the Cetus 
field, along with an analysis of X-ray source counts, stacked X-ray spectrum, 
and optical identifications. 
A total of 188 X-ray sources were detected: 174 in the 0.5 -- 7.0
keV band, 154 in the 0.5 -- 2.0 keV band, and 113 in the 2.0 -- 7.0 
keV band.
The X-ray source counts were derived and compared with 
LALA Bo\"{o}tes field (172 ks exposure).
Interestingly, we find consistent hard band X-ray source density,
but 36$\pm12$\% higher soft band X-ray source density in Cetus field.
The weighted stacked spectrum of the detected X-ray sources can be fitted by
a powerlaw with photon index $\Gamma$ = 1.55.
Based on the weighted stacked spectrum, we find that the 
resolved fraction of the X-ray background
drops from 72$\pm$1\% at 0.5 -- 1.0 keV to 63$\pm$4\% at 6.0 -- 8.0 keV.
The unresolved spectrum can be fitted by a powerlaw over the range 
0.5 -- 7 keV, with a photon index $\Gamma$ = 1.22.
We also present optical counterparts for 154 of the X-ray sources,
down to a limiting magnitude of $r$' = 25.9 (Vega), using a deep $r$' 
band image obtained with the MMT.
\end{abstract}


\altaffiltext{1}{Optical Observations reported here were obtained at the MMT Observatory,
a joint facility of the University of Arizona and the Smithsonian
Institution.}
\altaffiltext{2}{Center for Astrophysics, University of Science and Technology of China, Hefei, Anhui 230026, P. R. China; jxw@ustc.edu.cn.}
\altaffiltext{3}{School of Earth and Space Exploration, Arizona State University, Tempe, AZ 85287}
\altaffiltext{4}{Department of Physics, Arizona State University, Tempe, AZ 85287}
\altaffiltext{5}{Department of Physics and Astronomy, Johns Hopkins University, Baltimore, MD, 21218}
\section {Introduction}
The launch of the $Chandra$ $X$-$Ray$ $Observatory$ in 1999 has opened a new
era of X-ray astronomy, thanks to its superb spatial resolution and sensitivity (Weisskopf et al. 2002). 
In recent years, many deep $Chandra$ images
of the extragalactic sky have been obtained, including the 2~Ms $Chandra$ Deep 
Field North (CDF-N, e.g., Brandt et al.\ 2001; Alexander et al.\ 2003),
1~Ms $Chandra$ Deep Field South (CDF-S, Giacconi et al.\ 2002; 
Rosati et al.\ 2002), and many others at moderate depths (Stern et 
al.\ 2002; Yang et al.\ 2003; Mushotzky et al.\ 2000; 
Manners et al.\ 2003; Wang et al.\ 2004a, 2004b; 
see Brandt \& 
Hasinger 2005 for a full list of deep extragalactic $Chandra$ exposures).
The European Space Agency's X-ray Multi-Mirror 
Mission-Newton (XMM-Newton; Jansen et al. 2001) has provided 
a comparable number of surveys (again listed by Brandt \& Hasinger 2005).
These deep images alone resolved the majority 0.1 -- 10 keV X-ray background.
Worsley et al (2005) present a detailed study of the fraction of 
resolved X-ray background as a function of energy. Together with multi-band 
follow-up observations, these deep images provide large samples of X-ray 
sources, and enable studies of the formation and evolution of galaxies, 
clusters and groups of galaxies, large-scale structures, and super massive 
black holes (SMBH).

In this paper we present 
a new deep (174 ks) $Chandra$ ACIS exposure of the Large Area Lyman Alpha 
(LALA; Rhoads et al. 2000) survey's Cetus field. This is the second of two deep
$Chandra$ images of LALA fields. The previous one is an 172 ks ACIS
exposure of LALA Bo\"{o}tes field (Malhotra et al. 2003, Wang et al. 2004a).
The X-ray image presented here was originally obtained to investigate the 
X-ray properties of the LALA detected high redshift Ly$\alpha$ emitters
(Wang et al., 2004b). 
In this paper we present a full catalog of the detected X-ray sources,
along with an analysis of the resolved X-ray background, and 
optical identifications using deep broad band images obtained with
the 6.5m MMT + Megacam on Mt.~Hopkins, Arizona.

\section{X-ray Observations and Data Reduction}

A total of 176 ks of $Chandra$ ACIS exposure on the LALA Cetus field
was obtained in very faint (VFAINT) mode, composed of two individual
observations. The first observation, with 160.4 ks exposure, was taken on
UT 2002 June 13--14
({\it Chandra\/} Obs ID 4129).  The second observation,
with 15.4 ks exposure, was taken one day later (UT 2002 June 15, Obs ID 4402).
All four ACIS-I chips and the ACIS-S2, ACIS-S3 chips were used,
with the telescope aimpoint centered on the ACIS-I3 chip
for each exposure. The aimpoints of both exposures were
R.A. =  02:04:44.081, decl. = --05:05:17.36 (J2000.0).
Due to their large off-axis angle during the observations,
the ACIS-S chips have poorer spatial
resolution and effective area than the ACIS-I chips.
We have therefore chosen to ignore the ACIS-S chip data in this work.
We reduced the data using the packages CIAO 3.3 and CALDB 3.2.2 
(see http://asc.harvard.edu/ciao). The
level 1 data were reprocessed to clean the ACIS particle background
for the VFAINT mode observations, and filtered to include
only the standard event grades 0, 2, 3, 4, and 6. All bad pixels and columns
were also removed.  We excluded high background time intervals
from level 2 files, leaving a net exposure time of 174 ks.
We extracted three images 
from the combined event file: a soft image (0.5 -- 2.0 keV),
a hard image (2.0 -- 7.0 keV) and a total image (0.5 -- 7.0 keV). 
We cut the hard and total bands at 7 keV, because the effective area of
{\it Chandra\/} decreases and the particle/cosmic ray background rises 
above this energy, resulting in very inefficient detection of sky and source
photons. 

\section{Source Detection and Catalog}

We ran WAVDETECT (Freeman et al. 2002) on the extracted X-ray images
using a probability threshold of 1 $\times$ 10$^{-7}$ (corresponding
to 0.5 false sources expected per image). We used wavelet scales of
1, $\sqrt{2}$, 2, 2$\sqrt{2}$, 4, 4$\sqrt{2}$, 8, 8$\sqrt{2}$, 16, 
16$\sqrt{2}$, and 32 pixels (where 1 pixel = 0.492\arcsec).
A total of 188 X-ray sources were detected: 174 in the total band
(0.5 -- 7.0 keV), 154 in the soft band (0.5 -- 2.0 keV), and 113
in the hard band (2.0 -- 7.0 keV)\footnote{Changing the WAVDETECT probability
threshold to $1\times 10^{-6}$ increases the total number of detected
X-ray sources to 213.  However, this would increase 
the expected number of false detections to $\sim$ 5 
per image, and we prefer to publish a conservative catalog with high
reliability.}. The catalog of the
detected sources is presented in Table 1.

The source ID, IAU name, right ascension and declination are presented
in Table 1. The IAU name for the sources is CXOLALA2 JHHMMSS.s+DDMMSS.
Whenever available, source coordinates derived in the soft band image, 
which has the best spatial resolution among the three bands, are presented.
For sources not detected in the soft band, total (with a higher priority) 
or hard band positions are used as substitutions.
Column 5 in table 1 gives the 3 $\sigma$ uncertainties of the centroid 
positions directly 
given by WAVDETECT. 
We perform circular aperture photometry to measure the net counts
in each band for all the 188 detected sources (Column 6-8).
For each source, we defined a source region which is a circle
centered at the source position (column 3 and 4 in Table 1), with 
radius R$_s$ set to the 95\% encircled-energy radius of $Chandra$
ACIS PSF at 1.5 keV. 
R$_s$ varies in the range of 2$\arcsec$ to 15$\arcsec$ from the
center to the edge of the field.
Source photons were then extracted from
the circles, and the local background was extracted from an 
annulus with outer radius of 2.4 $\times$ R$_s$ and inner radius
of 1.2 $\times$ R$_s$, after masking out nearby sources.
The aperture correction ($\times$ 1/0.95) was applied to the
source counts.  
While this correction is based on the PSF at 1.5 keV, and could be refined
to account for the variation of PSF with photon energy and the spectral
slopes of sources, such higher order corrections would have a negligible
impact on our final count rates.
The derived
net counts and 1 $\sigma$ Poisson uncertainties\footnote{We use Bayesian method to get 
the net counts and 1 $\sigma$ Poisson uncertainties (e.g., Kraft et al, 1991).}
are given for each source in each band.

\begin{figure}
\plotone{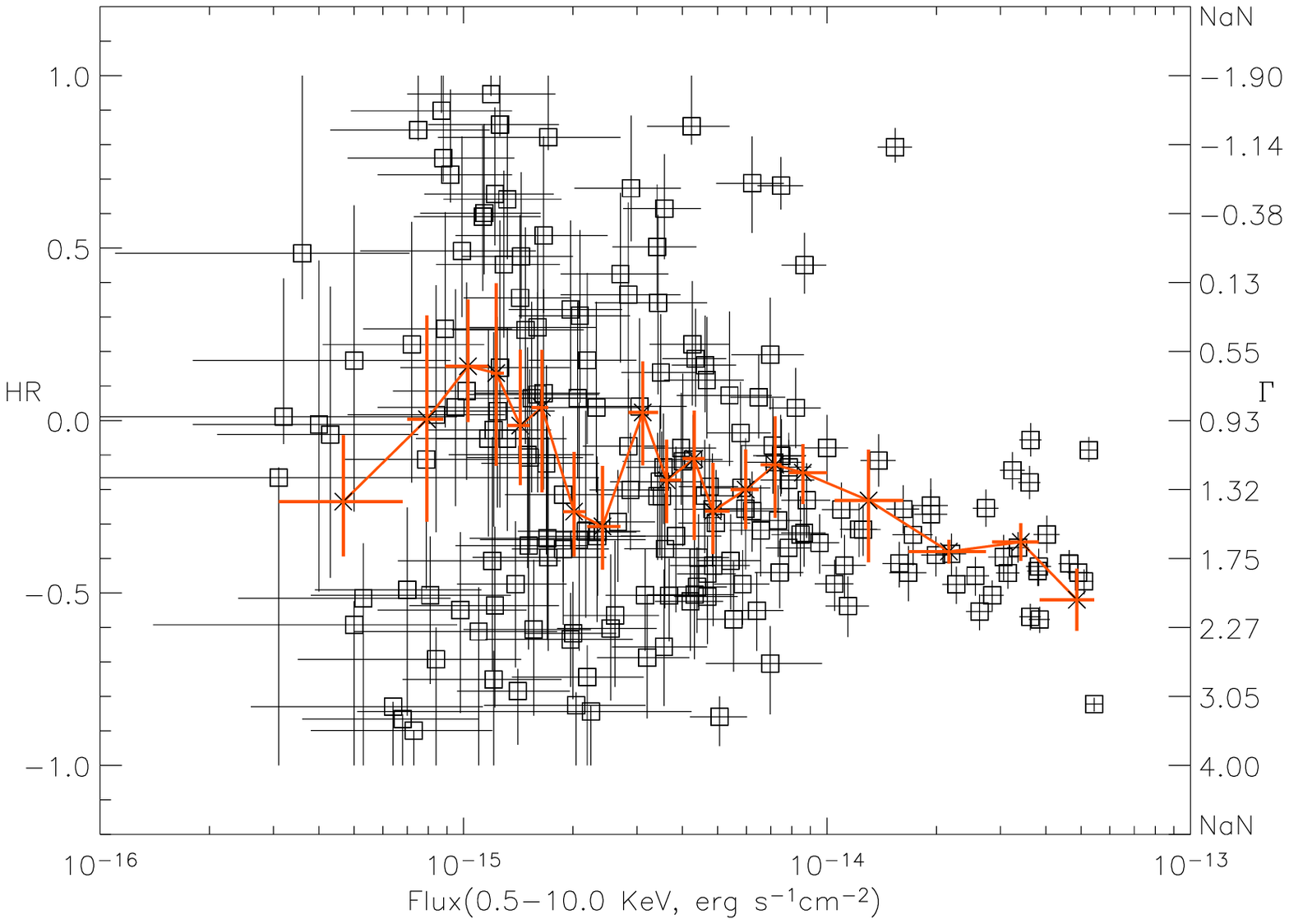}
\caption{Hardness ratios HR = $(H-S)/(H+S)$ (left ordinate axis) of X-ray sources vs their
full band (0.5 -- 10.0 keV) X-ray fluxes. The photon indices $\Gamma$ of
the power-law spectra which could reproduce the observed hardness
ratios are given along the right ordinate (see text for details). 
The line connects the average hardness ratios in different flux bins (10 sources
per bin).
}
\label{hr}
\end{figure}

In column 9 we mark the sources
with ``T'', ``S'' and ``H'' for detection in total, soft
and hard band respectively. Multiple letters are used for
sources detected in more than one band. 
In column 10 we present the hardness ratios. Here we calculate the hardness
ratios with a Bayesian approach which models the detected counts
as a Poisson distribution and can give reliable errorbars
for both low and high count sources (van Dyk et al. 2004; Park et al. 2006). 
The hardness ratios versus 0.5 -- 10.0 keV band X-ray fluxes for the detected
X-ray sources are plotted in Fig. \ref{hr}.
Assuming a power-law spectrum with
the Galactic HI column density (2.2 $\times$ 10$^{20}$ cm$^{-2}$),
the observed hardness ratio can be converted to the photon index $\Gamma$ of the
spectrum, which is also presented in the figure.
An increasing proportion of hard sources is seen at the fainter
flux levels (an effect seen also in earlier surveys).  Most of
these are believed to be obscured AGNs.

\begin{figure}
\plotone{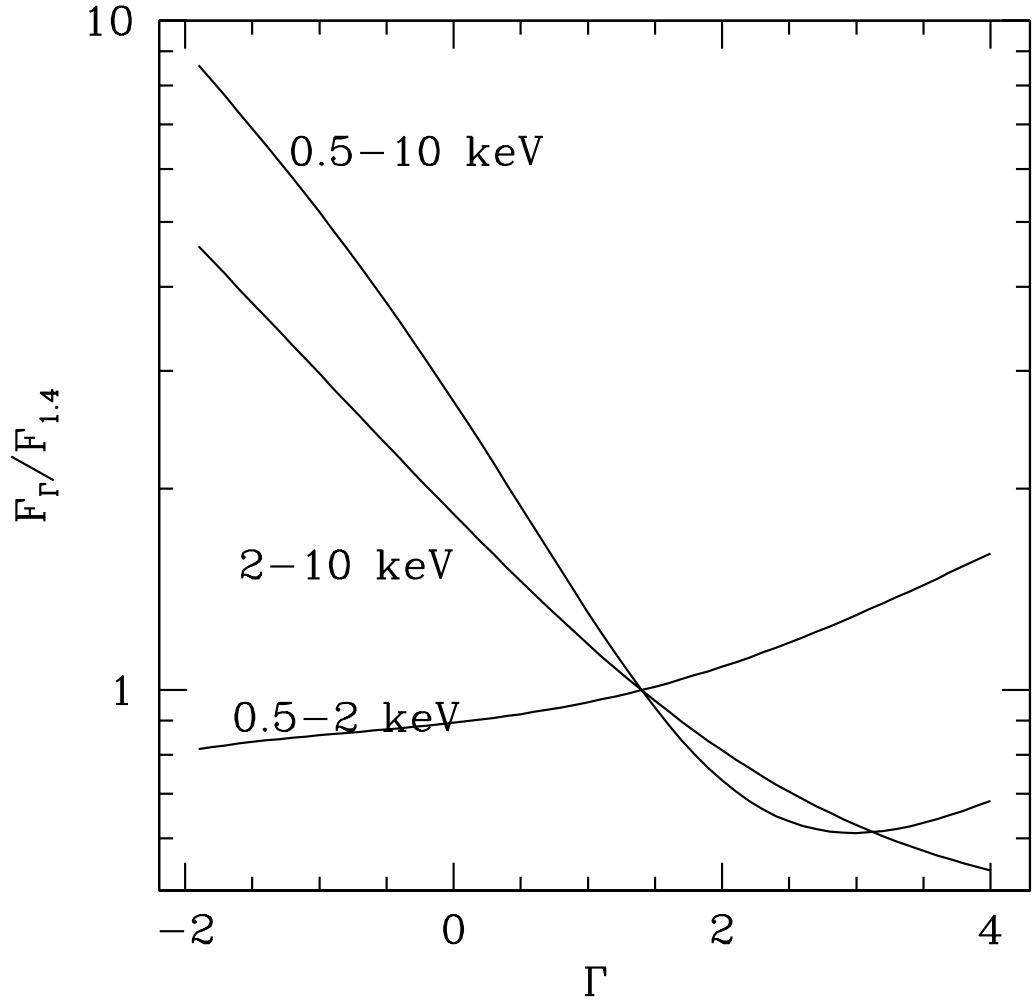}
\caption{Conversion factors to calculate three band fluxes from
count rates, assuming a
power-law spectrum with photon index different from $\Gamma$ = 1.4. 
}
\label{cf}
\end{figure}

In column 11-13 we present X-ray fluxes (Galactic absorption corrected, 
$N_H$ = 2.2 $\times$ 10$^{20}$ cm$^{-2}$, Dickey \& Lockman 1990) in 
three bands. 
We assumed a power-law spectrum, absorbed by the Galactic column 
density, to calculate the conversion factors from net counts to X-ray fluxes.
We used a photon index of $\Gamma$ = 1.4, as previously 
used by Giacconi et al. (2002), Stern et al. (2002) and Wang et al. (2004a). 
Net count rates were calculated by
dividing the net counts in column 6-8 by the effective exposure
time at each source position in each band.  We then 
converted these into X-ray fluxes in 0.5 -- 10.0 keV, 0.5 -- 2.0
keV, and 2.0 -- 10.0 keV bands respectively. 
The conversion factors adopted are 1.35 $\times$ 10$^{-11}$ ergs cm$^{-2}$
count$^{-1}$ for the total band; 
5.66 $\times$ 10$^{-12}$ ergs cm$^{-2}$ count$^{-1}$ for the soft band;
and 2.65 $\times$ 10$^{-11}$ ergs cm$^{-2}$ count$^{-1}$ for the hard band.
As we have pointed in Wang et al. (2004a), the total band 
flux differs from the sum of the soft and hard band fluxes if the actual 
photon index differs from 1.4. In Fig. \ref{cf} we plot the conversion factors
required assuming a power-law spectrum with different photon index $\Gamma$.
It is obvious in the figure that the conversion factors in hard 
and total band are quite sensitive to the assumed $\Gamma$.
We thus need to be cautious while using the fluxes obtained assuming a
constant $\Gamma$.

\section {X-ray source counts}
\begin{figure}
\plotone{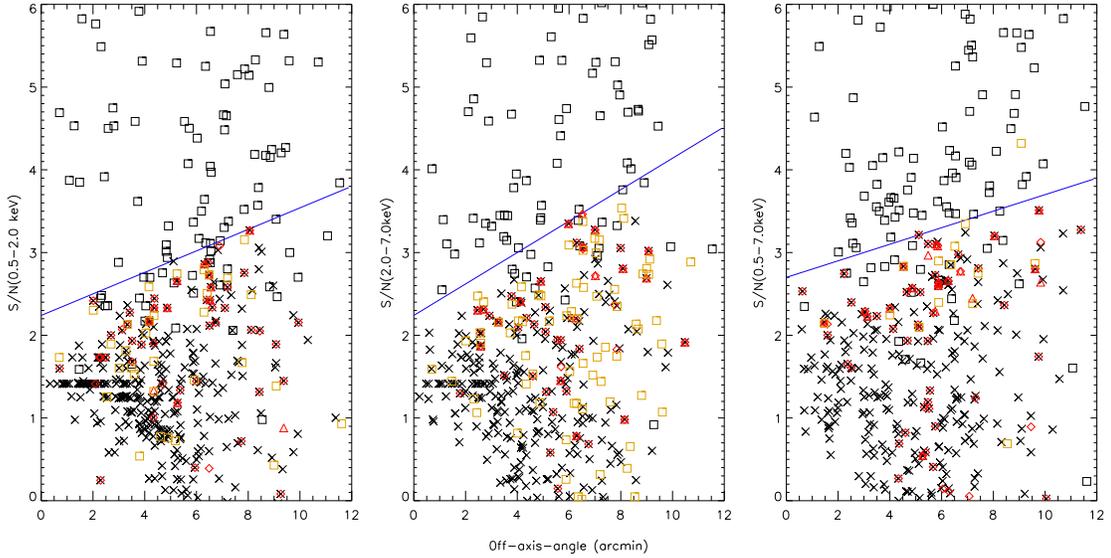}
\caption{Soft (0.5 -- 2.0 keV), hard (2.0 -- 7.0 keV), and total 
(0.5 -- 7.0 keV) band signal-to-noise ratio vs off-axis angle. 
The blue lines are the threshold we chose to build complete samples 
for our LogN-LogS calculation.
The signal-to-noise ratios are derived from X-ray photometry (see text
for details).  Thus, for each source, we can give the SN ratio in each band, 
even it is not detected by WAVDETECT in that band.
In these figures, the open black squares are sources detected in corresponding 
band (using WAVDETECT threshold of 10$^{-7}$), and other cataloged sources 
which are not detected in that band are plotted as yellow squares.
We also plot additional sources detected by WAVDETECT but with lower thresholds
(black crosses, red diamonds and red triangles with WAVDETECT thresholds of
10$^{-4}$, 10$^{-5}$ and 10$^{-6}$ respectively).
}
\label{cutoff}
\end{figure}

Since $Chandra$ ACIS does not have uniform sensitivity across the field
of view, our survey's sky coverage is a function of source flux.
(That is, brighter sources can be detected with larger sky area.)
To calculate the sky coverage, we first determine the formal signal
to noise ratio, 
\begin{equation}
S/N = (NET_{cts})/\sqrt{(NET_{cts} + BG_{cts})},
\end{equation}
that is required to ensure a high confidence detection in WAVDETECT,
as a function of off-axis angle in the field.  To do this,
we present in figure~\ref{cutoff} the soft, 
hard, and total band S/N ratio versus off-axis angle $\theta$ 
for the X-ray sources detected in each band. 
In each panel, nondetections are displayed with different symbols. 
(``Nondetections'' here means sources not detected in that 
band, along with additional sources detected in that band but using
a less significant WAVDETECT  probability 
threshold\footnote{We use different probability threshold of 
WAVDETECT, $1\times 10^{-6}$, $1\times 10^{-5}$ and $1\times 10^{-4}$ , 
to extract additional sources.})

\begin{figure}
\plotone{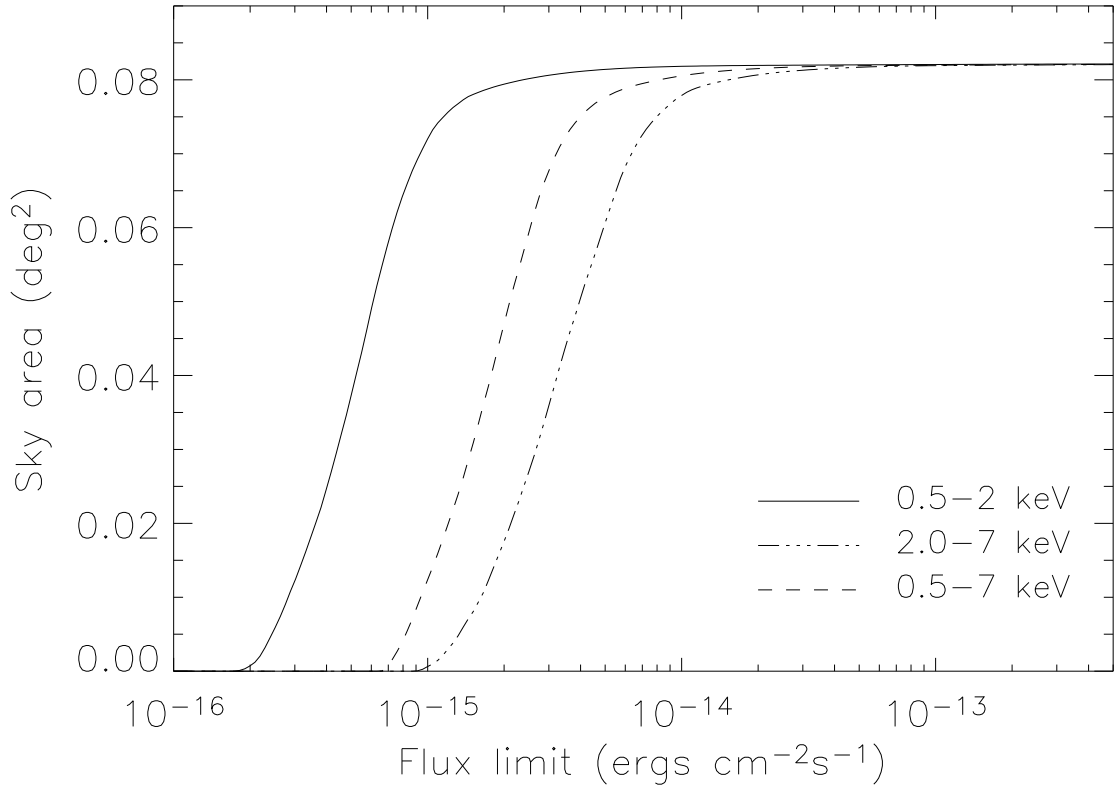}
\caption{Sky coverage in soft (0.5 -- 2.0 keV), hard (2.0 -- 7.0 keV) 
and total (0.5 -- 7.0 keV) bands, as a function of flux limit.}
\label{sc}
\end{figure}

\begin{figure}
\plotone{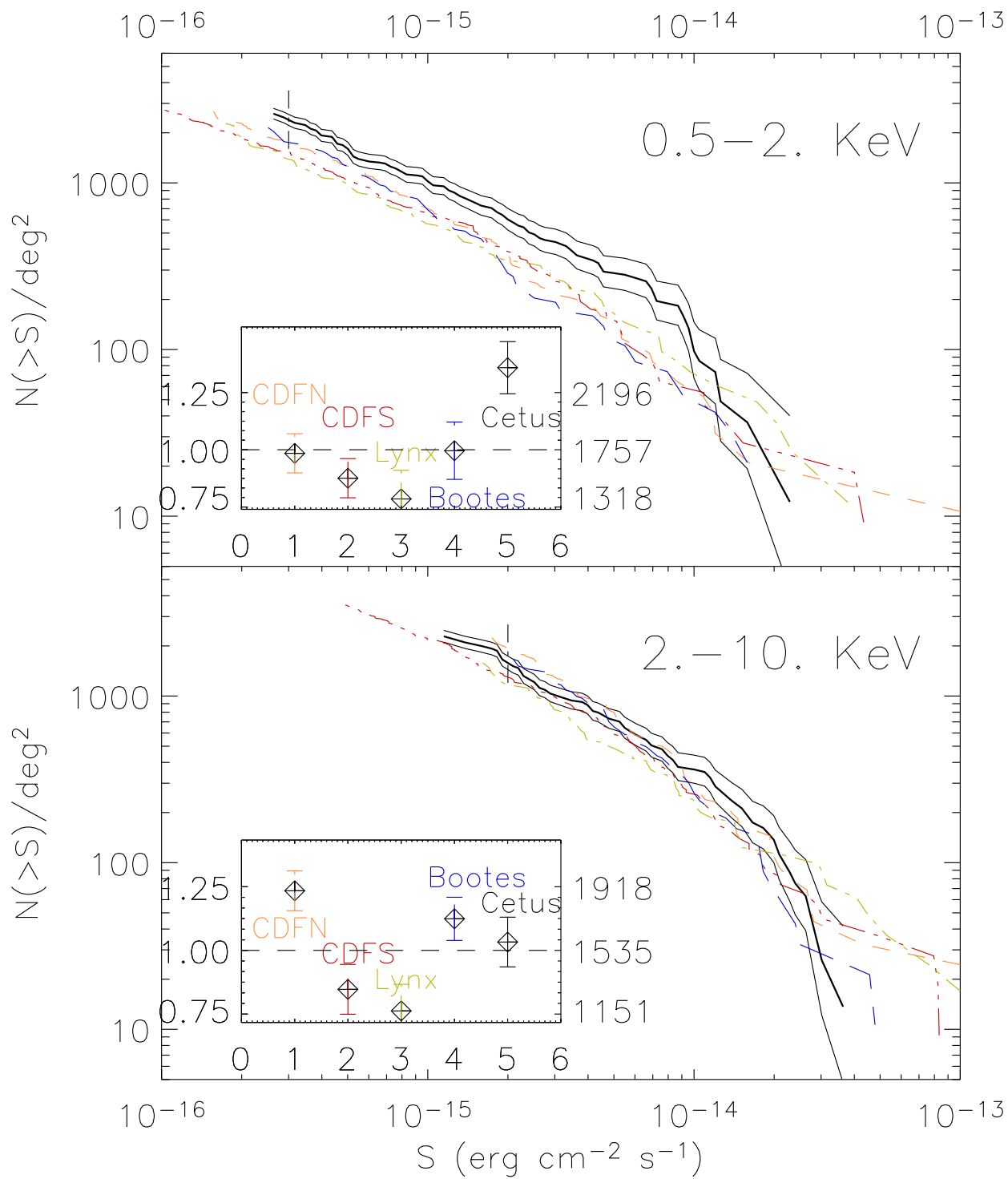}
\caption{Integral source counts (LogN-LogS) in the soft (0.5 -- 2.0 keV) 
and hard (2.0 -- 10.0 keV) bands from $Chandra$ observations of LALA 174 ks  
Cetus field, plotted as thick solid lines with two
additional thin solid lines enclosing 1$\sigma$ Poisson uncertainties.
The LogN-LogS from the LALA 172 ks Bootes (dashed line in blue), 2 Ms 
CDF-N (long-dashed line in orange), 
1 Ms CDF-S (dash-dotted line in red) and Lynx (dotted line in yellow) fields 
are also plotted. The inserts show the X-ray source densities and 1$\sigma$ 
uncertainties from the five fields at the faint end of our 174 ks $Chandra$
exposure (3.0$\times10^{-16}$\fluxunit\ in the
0.5 -- 2.0 keV band, and 2.0$\times10^{-15}$\fluxunit\ in the
2.0 -- 10.0 keV band). The average source densities from the five fields
(dashed lines) are also shown in the  inserts.
}
\label{lognlogs}
\end{figure}

It's clear from the figure that the minimum S/N required to ensure
a high confidence detection increases with the off-axis angle.
Thus, instead of using a constant S/N limit across the field,
we use a variable S/N limit dependent on the off-axis angle.
Three dashed lines ( $S/N  = A + B \times \theta$ respectively) are 
shown in  
fig.~\ref{cutoff}, with the parameters A and B chosen by visual 
inspection to include maximum number of detected sources in each band
but at most one nondetection above the lines.  Although there are 
detected sources below the threshold lines, they are mixed up with 
nondetections, i.e., the detection is incomplete below the lines. 
In this paper, we use only these sources with 
$S/N > A + B\times \theta$ to calculate LogN-LogS\footnote{A = 2.24, 2.24, 2.7 and
B = 0.13, 0.19 0.10 for the soft, hard and tot band respectively.}.
We then use our derived cutoff as a function of angle to determine 
the sky coverages as a function of flux.  The resulting area-flux
curves are shown in Fig.~\ref{sc}, and the resulting $\log{N}$-$\log{S}$
relations for both soft and hard bands in Fig.~\ref{lognlogs}.
We measured the slope of the  $\log{N}$-$\log{S}$ with a maximum
likelihood power law fit in each band.  For the 0.5 -- 2.0 keV band we find
\begin{equation}
N (> S) = 542^{+18}_{-18} \times  \left({ S \over {2. \times 10^{-15}\ {\rm ergs}\ {\rm cm^{-2}}\ {\rm s^{-1}} } }\right)^{-0.806^{+0.024}_{-0.024}}
\end{equation}
and for the 2.0 -- 10.0 keV band
\begin{equation}
N (> S) = 1577^{+52}_{-53} \times \left({ S \over {2. \times 10^{-15}\ {\rm ergs}\ {\rm cm^{-2}}\ {\rm s^{-1}} } }\right)^{-1.033^{+0.036}_{-0.035}}
\end{equation}
The error bars here are 1$\sigma$ errors derived from 
$\Delta\chi^2_{\nu=2}=2.30$ (e.g., Lampton et al. 1976).

For comparison, we re-reduced the 172 ks Chandra exposure on LALA Bo\"{o}tes 
field following the same procedures described above, and obtained the source
counts. This minimized the possible bias due to the difference between 
calibrations and data reductions. In the Bo\"{o}tes filed, we detected 167 
sources in the 
0.5 -- 7.0 keV band, 139 in the 0.5 -- 2.0 keV band, and 118 in the 2.0 -- 7.0
keV band. Note the numbers are slightly different from those published
in Wang et al (2004a) due to the different version of calibration files and
source detections procedures.
Comparing with the Bo\"{o}tes field, in the Cetus field 
we detect fewer sources in the hard band (113 vs 118), but more sources in
the soft band (154 vs 139).
Since the source numbers in the soft and hard band are not independent 
quantities, 
the soft band over density in the Cetus field is much 
more significant than it appears in the number ratio 154 vs 139.
A rough estimation can be made based the facts that soft band detected
sources outnumbers hard band ones by 41 in the Cetus field and 21 in the
Bo\"{o}tes field. This (41 vs 21) corresponds to a 2.8$\sigma$ excess
in the Cetus field. Note such calculation is only valid if all of the 
hard band detected sources are also detected in the soft band.
For our case in this paper, the accurate significant level should be close 
to 2.8$\sigma$ since most (82\%) of the hard band detected sources in both 
fields show up in the soft band. 

Note the actual source density depends on the sky coverage and also the
Galactic absorption which affects the observed soft band count rates.
We point out the soft band excess in the Cetus field
can not be attributed to these effects, since it has even slightly smaller 
sky coverage and larger Galactic absorption comparing with the Bo\"{o}tes 
field. In the plot of LogN-LogS (see Fig. \ref{lognlogs}) where 
these effects have been corrected, it 
is also clear that while two fields are consistent
in the hard band source density, the Cetus field has
36$\pm12$\% higher source density
(substantially above the Poisson noise) in the soft band
at 2.0 $\times$ 10$^{-15}$ erg cm$^{-2}$ s$^{-1}$. 
Following Wang et al. (2004a), we also plot LogN-LogS from different fields
for comparison (CDF-N, Brandt et al. 2001; CDF-S, Rosati et al. 2002; 
Lynx, Stern et al. 2002).
We note that the soft band excess in the Cetus field is also obvious 
(36$\pm12$\%) comparing with the mean of the five fields.

\section {The resolved X-ray background and the stacked X-ray spectrum}

What fraction of the hard X-ray background is resolved by our
deep 174 ks $Chandra$ imaging? We try to find the answer with two different
approaches.
We first calculate the fraction from the LogN-LogS.
In the flux range of 1.15 -- 100 $\times$ 10$^{-15}$ \fluxunit, 
the integrated hard X-ray flux density in the 2.0 -- 10.0 keV band is 1.28 
$\times$ 10$^{-11}$ \fluxunit\ deg$^{-2}$. This corresponds to 
57\% of the 2. -- 10.0 keV band X-ray background (De Luca \& Molendi 2004). 
Note that there is no 
source brighter than 1.0 $\times$ 10$^{-13}$ \fluxunit\ in the Cetus field.
By combining with the wide area $ASCA$ Large Sky Survey (with sky
coverage of $\sim$ 7 deg$^2$, Ueda et al. 1999) which resolves
0.46 $\times$ 10$^{-11}$ \fluxunit\ deg$^{-2}$ down to a flux limit of 
1.0 $\times$ 10$^{-13}$ \fluxunit\ in the 2 -- 10 keV band, we obtained
a higher fraction of 77.6\%.

\begin{figure}
\figurenum{6}
\plotone{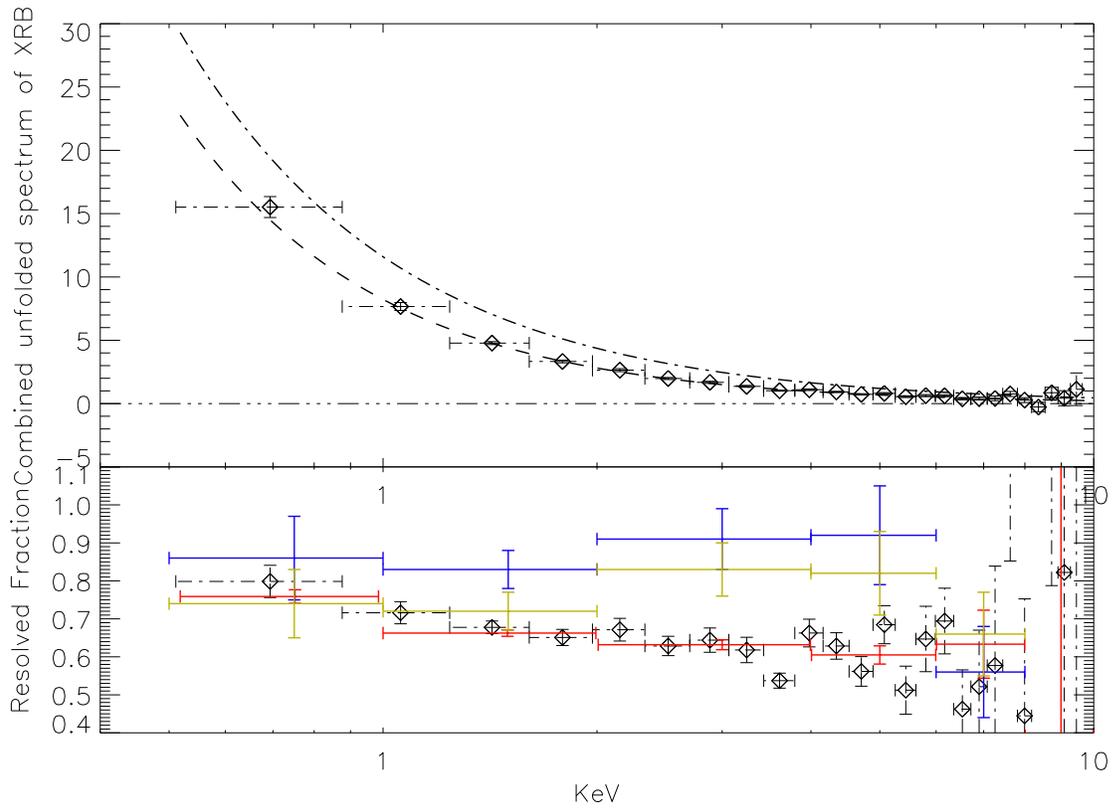}
\caption{Stacked X-ray spectrum (top) and resolved fraction (bottom). 
The stacked spectrum is binned for display purpose, plotted with 1$\sigma$ 
errorbars, and fitted with a powerlaw model (plotted as dashed line). The XRB 
model is presented as dot-dashed line.
Blue and yellow data points are resolved fractions in different bands in CDF-N 
and CDF-S obtained by Worsley et al. (2005). Red data points are values in the
Cetus field in corresponding bands.
}
\label{ufs}
\end{figure}

Recently attempts have been made to measure the resolved fraction of XRB as
a function of energy in deep Chandra/XMM surveys. By stacking X-ray sources
in Chandra Deep Fields and XMM Lockman Hole in narrower photometric bands,
Worsley et al. (2004; 2005) have shown that the resolved fraction of
XRB drops from 80-100\% at $<$ 2 keV to $\sim$ 50\% at above 8 keV.
The fraction as a function of energy can be used to estimate the spectral 
shape of the unresolved XRB and the population of sources contributing it.
In this paper we adopt a different approach to estimate the intensity
spectrum of the resolved XRB in the LALA Cetus field.
The resolved XRB intensity spectrum was obtained by summing the unfolded spectrum for each
detected X-ray source divided by its sky coverage. In \S4 we have obtained the
soft, hard and total band sky coverage for each source.  We use the 
largest of these three areas for each source when constructing our
composite spectrum.  We exclude sources with sky coverage $<$ 0.003 
degree$^2$, to avoid large uncertainties due to small sky coverage.
The unfolded spectra were obtained assuming a powerlaw with
$\Gamma$ = 1.4 absorbed by Galactic hydrogen column density. Adopting
slightly different $\Gamma$ as 1.2 or 1.6 does not change our results below.
The stacked X-ray spectral intensity is presented in Fig. \ref{ufs}, which
is best-fitted by a powerlaw with photon index $\Gamma$ = 1.55 $\pm$ 0.02
and normalization of 
8.23 $\pm$ 0.07 photons cm$^{-2}$s$^{-1}$sr$^{-1}$keV$^{-1}$.

We adopt the XRB model spectrum used by Worsley et al. (2005), which
is a power-law with a spectral slope of $\Gamma = 1.41$ in the 
1 -- 8 keV band (De Luca \& Molendi 2004).  Below 1 keV, we use 
the measurement at
0.25 keV from Roberts \& Warwick (2001), which yields a spectral slope of
$\Gamma$ = 1.6.
We thus see that the resolved fraction of the X-ray background in Cetus field
can be modeled as 0.72 $\times$ $E^{-0.14}$, decreasing from 72\% at 1 keV to
54\% at 8 keV. There are very large uncertainties above 8 keV, due to the
limited photons in that band.
We did not attempt to make corrections for very bright sources which are
absent in such a small field.
We also plot the resolved fraction of XRB in CDF-N and CDF-S obtained by
Worsley et al. (2005). In the 0.5-1 keV band, the resolved fraction in the
Cetus field is comparable with those of CDFS. This is mainly due to the higher
source density in the Cetus field as we found in \S4.
The fraction decreases from 1 keV to 8 keV and is obviously smaller
than CDFs, because of the less exposure time in the Cetus field.
We note that to avoid the bias caused by cosmic variance,
accurate measurements of the resolved and unresolved spectra require combining 
deep surveys with shallow wide area surveys.

\section {Optical Identifications}
Deep broadband images on LALA Cetus field were obtained in four Sloan Digital
Sky Survey (SDSS) filters ($g$',$r$',$i$',and $z$') using the Megacam
instrument (Mcleod et al.1998) at the MMT. 
The total exposure time in the $g$',$r$',$i$',and $z$' 4.33, 3.50, 4.78
and 5.33 hours respectively. The details of the observations and data reduction
have been presented in Finkelstein et al. (2007), who used the deep broadband
images to study the ages and masses of Ly$\alpha$ emitters at $z$ $\sim$ 4.5. 
The 5 $\sigma$ limiting magnitudes (2.32\arcsec diameter aperture) obtained 
are 26.38, 25.64, 25.13 and 24.1 (Vega) respectively in four bands.

We use the MMT $r$' band to identify our X-ray 
sources. We find the astrometry of the X-ray image and the MMT images are
well matched, with an average shift less than 0.1\arcsec.
We use circles with radius equals the root sum square of the 3$\sigma$ 
positional uncertainties from WAVDETECT for X-ray sources (see table 1) 
and 1\arcsec.
Here 1\arcsec\ stands for the uncertainty in the absolute X-ray astrometry.
Optical counterparts in $r$' band are found for 158 X-ray sources. We find
two possible counterparts each for three of these X-ray sources. In table 1,
we only present the one which is closer to the center of the search circle.
Two of the X-ray sources are located outside the MMT image.
There are also 8 sources which are too close to nearby bright sources, making
it impossible to identify them in the optical band image and to provide 
reliable upper limits to their $r$' band magnitudes. For the remaining
20 X-ray sources without $r'$ band counterparts, we tabulate the 
5$\sigma$ $r$' band upper limits, i.e. 25.86\footnote{This is the mean Vega 
magnitude of SExtractor detected sources with signal to noise ratio 
around 5, and thus differs slightly from the 5 $\sigma$ limiting 
magnitude in a 2.32\arcsec diameter aperture that is 
reported above.}).

In column 14-16 in Table 1 we present the offsets of the detected optical 
counterparts from the X-ray source positions ($\Delta\alpha$ =
RA$_R$ - RA$_X$, $\Delta\delta$ = Dec$_R$ - Dec$_X$),
and the $r$' band AUTO magnitudes from SExtractor
(Kron-like elliptical aperture magnitudes, Bertin \& Arnouts 1996)
together with their 1 $\sigma$
uncertainties\footnote{The uncertainties of the
magnitudes are direct output from SExtractor, without including
the uncertainty of the $r$' band zeropoint.}. For sources which are not
detected in $r$' band, 5$\sigma$ upper limits of the magnitudes
are given when available.

Among the 20 X-ray sources without $r$' band counterparts, four have 
counterparts in the $g$' band MMT image. The remaining 16 sources, 
which are not detected either in the $g$' or $r$' band, are 
possible candidates for $z > 5$ quasars,
since such sources would be invisible in the $g$' or $r$' band because of the absorption from the Lyman transitions of hydrogen along our
line of sight.   However, 8 of these 16 sources have X-ray hardness ratio
$HR > 0.0$, indicating strong X-ray obscuration. This suggests that they are
X-ray obscured AGNs at low to at most intermediate redshift, because
even heavily obscured AGNs at $z>5$ would appear soft in {\it Chandra}
images due to redshifting of the X-ray spectrum (Wang et al.\ 2004c).
This reduce the number of candidate $z>5$ quasars to 8.
A program to obtain the optical spectra of these X-ray sources using
the Inamori-Magellan Areal Camera and Spectrograph (IMACS) on the 
Magellan 6.5m telescopes is under way.

\section {Conclusions}
We present X-ray sources detected in 174 ks of $Chandra$ ACIS integration
on the LALA Cetus field.
A total of 188 X-ray sources were detected: 174 in the total band
(0.5 -- 7.0 keV), 154 in the soft band (0.5 -- 2.0 keV), and 113
in the hard band (2.0 -- 7.0 keV). The detection near the aimpoint
is down to a flux limit of 2.5 
$\times$ 10$^{-16}$ \fluxunit\ in the soft (0.5 -- 2.0 keV) band, and 1.15 
$\times$ 10$^{-15}$ \fluxunit\ in the hard (2.0 -- 10.0 keV) band.
Comparing with LALA Bo\"{o}tes field, we obtain comparable X-ray source
density in the hard band, but 36$\pm12$\% higher source density in the 
soft band. 
We study the resolved X-ray background intensity by stacking the X-ray spectra
of detected X-ray sources, weighted by sky coverage for each source. We find 
that the resolved fraction of the X-ray background
drops from 72$\pm$1\% at 0.5 -- 1.0 keV to 63$\pm$4\% at 6.0 -- 8.0 keV,
 consistent with previous works.
The unresolved spectrum can be fitted by a powerlaw at 0.5 -- 7 keV with
a photon index $\Gamma$ = 1.22$\pm$0.04. We also present
$r$' band optical counterparts for 158 of the X-ray sources.

\acknowledgements 
The work of JXW is supported by Chinese NSF through NSFC10473009, NSFC10533050 and the CAS ``Bai Ren'' project at University of Science and Technology of China.

\clearpage

\newpage

\begin{deluxetable}{r@{~~~}l@{}r@{~~~}r@{~~~}r@{~~~}r@{~~~}r@{~~~}c@{~~~}r@{~~~}r@{~~~}r@{~~~}c@{~~~}c@{~~~}l@{~~~}c@{~~~}c@{~~~}c@{~~~}c@{~~~}}
\rotate
\tabletypesize{\tiny}
\tablewidth{0pc}
\tablecaption{$Chandra$ Sources in LALA Cetus Field\label{common}}
\tablehead{
\colhead{ID} & \colhead{\space\space CXOLALA2\space\space\space\space\space\space\space\space\space RA(J2000)\space\space  Dec(J2000)} & \colhead{Err} & \colhead{Tot. Cts.} & \colhead{Soft Cts.} & \colhead{Hard Cts.} & \colhead{det} & \colhead{HR} & \colhead{S$_{0.5-10}^e$} & \colhead{S$_{0.5-2}^e$} & \colhead{S$_{2-10}^e$} & \colhead{$\Delta\alpha(\arcsec)$} & \colhead{$\Delta\delta(\arcsec)$} &\colhead{$r$}  }

\startdata
1 & J020457.9-05057.4\space\space\space02:04:57.9\space\space\space-05:05:7.4 & 0.25\arcsec & $ 97.2^{+14.8}_{-13.5 }$ & $ 55.5^{+11.2}_{-9.9 }$ & $ 41.7^{+10.0}_{-8.7 }$ & TSH & $ -0.14^{+0.10}_{-0.10 }$ & 7.85 & 1.88 & 6.55 	& -0.5 & -0.4 & 21.79$\pm$0.01\\
2 & J020456.9-050337.3\space\space\space02:04:56.9\space\space\space-05:03:37.3 & 0.15\arcsec & $ 437.9^{+30.3}_{-29.2 }$ & $ 300.1^{+25.2}_{-24.1 }$ & $ 137.8^{+17.3}_{-16.0 }$ & TSH & $ -0.37^{+0.04}_{-0.04 }$ & 33.55 & 9.66 & 20.58 	& -0.0 & 0.1 & 19.52$\pm$0.00\\
3 & J020452.9-050828.9\space\space\space02:04:52.9\space\space\space-05:08:28.9 & 0.22\arcsec & $ 140.7^{+17.5}_{-16.3 }$ & $ 88.8^{+14.0}_{-12.8 }$ & $ 51.8^{+10.9}_{-9.7 }$ & TSH & $ -0.26^{+0.08}_{-0.08 }$ & 10.95 & 2.92 & 7.82 	& -0.1 & 0.2 & 24.25$\pm$0.07\\
4 & J020450.2-050350.6\space\space\space02:04:50.2\space\space\space-05:03:50.6 & 0.26\arcsec & $ 56.6^{+11.4}_{-10.1 }$ & $ 34.1^{+9.0}_{-7.7 }$ & $ 22.5^{+7.5}_{-6.1 }$ & TSH & $ -0.20^{+0.12}_{-0.14 }$ & 5.82 & 1.47 & 4.52 	& 0.2 & 0.1 & 21.99$\pm$0.01\\
5 & J020448.2-050315.3\space\space\space02:04:48.2\space\space\space-05:03:15.3 & 0.09\arcsec & $ 694.8^{+37.7}_{-37.0 }$ & $ 377.9^{+28.0}_{-27.0 }$ & $ 316.9^{+25.9}_{-24.8 }$ & TSH & $ -0.09^{+0.04}_{-0.03 }$ & 52.52 & 11.98 & 46.74 	& 0.2 & 0.1 & 23.25$\pm$0.04\\
6 & J020448.1-050923.1\space\space\space02:04:48.1\space\space\space-05:09:23.1 & 0.22\arcsec & $ 159.1^{+18.6}_{-17.4 }$ & $ 105.1^{+15.2}_{-13.9 }$ & $ 54.0^{+11.2}_{-9.9 }$ & TSH & $ -0.32^{+0.07}_{-0.08 }$ & 12.57 & 3.51 & 8.24 	& -0.1 & -0.2 & 20.57$\pm$0.00\\
7 & J020447.6-050129.2\space\space\space02:04:47.6\space\space\space-05:01:29.2 & 0.29\arcsec & $ 38.7^{+9.8}_{-8.5 }$ & $ 29.6^{+8.5}_{-7.2 }$ & $ 9.1^{+5.4}_{-4.1 }$ & TSH & $ -0.51^{+0.13}_{-0.16 }$ & 3.16 & 1.02 & 1.45 	& 0.1 & 0.3 & 22.29$\pm$0.01\\
\enddata
\tablenotetext{a}{Sources with the searching circles seriously overlapped by 
nearby bright sources and no optical counterparts can be visually identified.
Reliable upper limits on their $r$' band magnitudes are thus not available}
\tablenotetext{b}{Two Sources which are not covered by the MMT $r$' band image}
\tablenotetext{c}{Sources with two counterparts identified within the searching
circles. Here we only present the one which is closer to the center of the circles.}
\tablenotetext{d}{Sources with optical counterparts visually identified. SExtractor failed to detect these counterparts due to dilution of nearby bright sources. For these counterparts, their $r$' band magnitudes are measured within a 1.5\arcsec radius circle.}
\tablenotetext{e}{in unit of 10$^{-15}$ erg cm$^{-2}$ s$^{-1}$.}
\end{deluxetable}
\end{document}